\documentstyle[12pt]{article}
\begin{document}
\title{CP Violation in $\gamma \gamma \rightarrow t \bar{t}$ within the
       Minimal Supersymmetric Standard Model}

\vspace{-3mm}
\author{{Zhou Mian-Lai$^{b}$, Ma Wen-Gan$^{a,b,c}$, Han Liang$^{b}$,}\\
        {Jiang Yi$^{b}$ and Zhou Hong$^{b}$}\\
{\small $^{a}$CCAST (World Laboratory), P.O.Box 8730, Beijing 100080, China} \\
{\small $^{b}$Modern Physics Department, University of Science and}\\
{\small Technology of China, Anhui 230027, China}  \\
{\small $^{c}$Institute of Theoretical Physics, Academia Sinica,
        P.O.Box 2735, Beijing 100080, China.}}

\date{}
\maketitle
\vspace{-8mm}

\begin{abstract}
The complete analysis of the CP violation in the process $\gamma \gamma
\rightarrow t \bar{t}$ in frame of the Minimal Supersymmetric Model(MSSM)
is presented. The CP-odd observables for describing the CP violating effects
in polarized and unpolarized photon collisions, are calculated. We
investigate the possible CP violation sources induced by the complex soft
breaking parameters and study the CP violating effects contributed by gluino,
neutralino and chargino sectors appearing in the loop diagrams. We find
that it is possible to observe the CP violating effects in top quark pair
production via polarized and unpolarized photon fusions by using optimal
observables and favorable parameters. \\
\end{abstract}

\vfill \eject


\noindent{\large\bf I. Introduction}

    Since the first discovery of CP violation in the kaon system over thirty
years ago \cite{kaon}, CP violating phenomena have been investigated and
discussed extensively by a lot of physicists. Various models have been
proposed to explain the CP violation observed in $K^{0}-\bar{K}^{0}$ mixing
\cite{kaonCP}. Although the Cabibbo-Kobayashi-Maskawa(CKM) \cite{CKM}
mechanism in the Standard Model(SM) can explain all the currently available
experimental data of CP violation, there is still some room for the
extended models of the SM, which can explain the CP violation phenomena
equivalently well. Moreover, in cosmology theory there is a so-called
baryon genesis problem of the universe \cite{baryon}, and the strength of
CP violation due to the CKM mechanism is not strong enough. It indicates
some new source(s) of CP violation is required. Therefore, the origin of
CP violation still remains a puzzle. The Standard Model accommodated with
the complex phase in CKM matrix gives the CP violating interactions, but
its predicted CP violating effects outside the K-, D-, and B-meson systems
are greatly suppressed at high energy scale and hence unobservably small
\cite{CPsmall}. That is to say, the experimental and theoretical discovery of
strong CP violation at large energy scale would probably reveal new physics
beyond the Standard Model, in which the CP violating effects might be
considerably enhanced, especially in the processes involving heavy quarks.
Thus, we shall concentrate on the observable effects of CP violation induced
by non-standard model interactions.

    Recently the evidence for the existence of the top quark has been found
experimentally by the CDF Collaboration with the top quark mass determined
being in the range of 170-200 GeV\cite{topmass}, which coincides with the
indirect determination from the precise data of electroweak experiments.
Because of the large mass of the top quark, it is believed
possible to probe the CP violation from the reactions involving top quarks.
Since then a lot of research has been carried on the CP violation in top
quark pair production. A. Bartl et.al. presented a complete analysis of
electric and weak dipole moment form factors of top quark with complex
supersymmetric parameters in \cite{Bartl}. References \cite{ee1} \cite{ee2}
\cite{ee3} and \cite{ee4} discussed the CP violation in the process $e^{+}
e^{-} \rightarrow t \bar{t}$, while the CP asymmetry in the top quark pair
production at hadron colliders was investigated in Ref.\cite{hardron1}
\cite{hardron2} \cite{hardron3} and \cite{hardron4}. It is known that the
process $\gamma \gamma \rightarrow t \bar{t}$ has more advantages in the
probing of CP violation because the production of top quark pairs via
photon-photon collision is much cleaner than at $pp$ or $p\bar{p}$ colliders,
and its production rate from back-scattering photons is much larger than that
from the direct $e^{+} e^{-} \rightarrow t \bar{t}$ production. At the Next
Linear Collider (NLC), a large number of top quark pairs are produced with
large statistical events \cite{rate}. On this aspect, Anlauf et.al. discussed
CP violation in a Higgs mediated $\gamma \gamma \rightarrow t \bar{t}$ process
and studied the triple-product correlations as well as other asymmetric
parameters \cite{anlauf}. Poulose et.al. also analyzed the electric dipole
moment of the top quark which leads to the CP violating asymmetries\cite{ee4}.
And Han Liang et.al. investigated the QCD corrections to the cross section and
the CP violating effects of $\gamma \gamma \rightarrow t \bar{t}$ with both
polarized and unpolarized initial particles \cite{han}.

    In this paper we study CP violation in the process $\gamma \gamma
\rightarrow t \bar{t}$ in the framework of the Minimal Supersymmetric
Standard Model(MSSM) \cite{MSSM}. It is the simplest case of the SUSY
model and the currently most favorite extension of the Standard Model.
In this model some strong CP violating interactions may be introduced,
which would enhance the CP violation effects greatly. As we know, in
the SM, the mixing of top quark with other generation is very small
because of the unitary of the CKM matrix, the corresponding CP violation
in this process would be negligibly small. However, in the MSSM, new
source(s) of CP violation can be induced by additional complex couplings
within one generation \cite{onegene}. This feature provides us with the
possibilities to investigate the strong CP violating phenomena in the
process $\gamma \gamma \rightarrow t \bar{t}$ within the MSSM.

    We organized the paper as follows. In Section II, we introduce two
CP-odd observables respectively for polarized and unpolarized photon
collisions, and analyze the CP asymmetries caused by the complex interactions
in the frame of the MSSM at one-loop level. Then in Section III,
the numerical calculation and discussion are presented. Finally, a short
summary is given. In Appendix we listed the explicit forms of the
transformation matrices U, V, and N.

\vskip 5mm
\noindent{\large\bf II. CP-odd Observables}

   In this paper, we denote the process as:
$$
\gamma(p_{3}, \lambda_{1})\gamma(p_{4}, \lambda_{2}) \rightarrow
              t(p_{1}, s_{1}) \bar{t}(p_{2}, s_{2}),
\eqno{(2.1)}
$$
where $p_{1}$, $p_{2}$, $p_{3}$ and $p_{4}$ represent the four-momenta
of the outgoing top quark pair and incoming photons respectively, whereas
$s_{1}$, $s_{2}$, $\lambda_{1}$ and $\lambda_{2}$ denote the spin
four-momenta of top quark pair and the polarizations of incoming photons,
respectively.

   The Feynman diagrams for the process (2.1) at the tree-level
are shown in Fig.1(a) and the one-loop diagrams which contribute to CP
violating effect in the frame of the MSSM are plotted in Fig.1(b)-(j).
The relevant Feynman rules can be found in Ref. \cite{MSSM}.
Including all the diagrams appearing in Fig.1, the renormalized amplitude
for $t\bar{t}$ pair production in $\gamma\gamma$ collision is shown as
\begin{eqnarray*}
 &&   M_{ren}(\lambda_{1},\lambda_{2},s_{1},s_{2}) =
      M_{tree}(\lambda_{1},\lambda_{2},s_{1},s_{2}) +
      M_{one-loop}^{CP}(\lambda_{1},\lambda_{2},s_{1},s_{2}) \\
&=& M_{tree}(\lambda_{1},\lambda_{2},s_{1},s_{2}) +
      M_{self-energy}^{CP}(\lambda_{1},\lambda_{2},s_{1},s_{2}) +
      M_{vertex}^{CP}(\lambda_{1},\lambda_{2},s_{1},s_{2}) \\
&& + M_{box}^{CP}(\lambda_{1},\lambda_{2},s_{1},s_{2}) +
      M_{quartic}^{CP}(\lambda_{1},\lambda_{2},s_{1},s_{2}) \\
&=& \epsilon_{\mu}(p_{3},\lambda_{1})\epsilon_{\nu}(p_{4},\lambda_{2})
      \bar{u}(p_{1},s_{1}) [
          f_{1}\gamma^{\mu}\gamma^{\nu}+
          f_{2}\gamma^{\nu}\gamma^{\mu}+
          f_{3}p_{1}^{\nu}\gamma^{\mu}+
          f_{4}p_{2}^{\nu}\gamma^{\mu} \\
&& + f_{5}p_{1}^{\mu}\gamma^{\nu} +
          f_{6}p_{2}^{\mu}\gamma^{\nu}+
          f_{7}p_{1}^{\mu}p_{1}^{\nu}+
          f_{8}p_{1}^{\mu}p_{2}^{\nu}+
          f_{9}p_{2}^{\mu}p_{1}^{\nu}+
          f_{10}p_{2}^{\mu}p_{2}^{\nu} \\
&& + f_{11}\rlap/p_{3}\gamma^{\mu}\gamma^{\nu}+
          f_{12}\rlap/p_{3}\gamma^{\nu}\gamma^{\mu} +
          f_{13}\rlap/p_{3}p_{1}^{\nu}\gamma^{\mu}+
          f_{14}\rlap/p_{3}p_{2}^{\nu}\gamma^{\mu}+
          f_{15}\rlap/p_{3}p_{1}^{\mu}\gamma^{\nu} \\
&& + f_{16}\rlap/p_{3}p_{2}^{\mu}\gamma^{\nu}+
          f_{17}\rlap/p_{3}p_{1}^{\mu}p_{1}^{\nu}+
          f_{18}\rlap/p_{3}p_{1}^{\mu}p_{2}^{\nu}+
          f_{19}\rlap/p_{3}p_{2}^{\mu}p_{1}^{\nu}+
          f_{20}\rlap/p_{3}p_{2}^{\mu}p_{2}^{\nu} \\
&& + f_{21}\gamma_{5}\gamma^{\mu}\gamma^{\nu}+
          f_{22}\gamma_{5}\gamma^{\nu}\gamma^{\mu}+
          f_{23}\gamma_{5}p_{1}^{\nu}\gamma^{\mu}+
          f_{24}\gamma_{5}p_{2}^{\nu}\gamma^{\mu}+
          f_{25}\gamma_{5}p_{1}^{\mu}\gamma^{\nu} \\
&& + f_{26}\gamma_{5}p_{2}^{\mu}\gamma^{\nu}+
          f_{27}\gamma_{5}p_{1}^{\mu}p_{1}^{\nu}+
          f_{28}\gamma_{5}p_{1}^{\mu}p_{2}^{\nu}+
          f_{29}\gamma_{5}p_{2}^{\mu}p_{1}^{\nu}+
          f_{30}\gamma_{5}p_{2}^{\mu}p_{2}^{\nu} \\
&& + f_{31}\gamma_{5}\rlap/p_{3}\gamma^{\mu}\gamma^{\nu}+
          f_{32}\gamma_{5}\rlap/p_{3}\gamma^{\nu}\gamma^{\mu}+
          f_{33}\gamma_{5}\rlap/p_{3}p_{1}^{\nu}\gamma^{\mu}+
          f_{34}\gamma_{5}\rlap/p_{3}p_{2}^{\nu}\gamma^{\mu} \\
&& + f_{35}\gamma_{5}\rlap/p_{3}p_{1}^{\mu}\gamma^{\nu}+
          f_{36}\gamma_{5}\rlap/p_{3}p_{2}^{\mu}\gamma^{\nu}+
          f_{37}\gamma_{5}\rlap/p_{3}p_{1}^{\mu}p_{1}^{\nu}+
          f_{38}\gamma_{5}\rlap/p_{3}p_{1}^{\mu}p_{2}^{\nu} \\
&& + f_{39}\gamma_{5}\rlap/p_{3}p_{2}^{\mu}p_{1}^{\nu}+
          f_{40}\gamma_{5}\rlap/p_{3}p_{2}^{\mu}p_{2}^{\nu}
                          ] v(p_{2},s_{2}).
~~~~~~~~~~~~~~(2.2)
\end{eqnarray*}
Our calculation shows that the form factor coefficients $f_{i}(i=1,2,...,20)$
turn out to have no effect on CP violation, but they contribute to the total
cross section at one-loop order. On the other hand, the other
$20$ form factor coefficients affect CP-odd observables, but do not appear
in the total cross section of the process when the polarizations of initial
states and spins of final states are summed up. It should be noticed that all
the terms in Eq.(2.2) involving $f_{i}(i=21,22,...,40)$ contain $\gamma_{5}$,
while the others which involve $f_{i}(i=1,2,...,20)$ do not.

   Concerning the CP violation in process (2.1), three types of couplings may
have contributions to CP asymmetry: gluino in the vertex $t\tilde{t}\tilde{g}$,
chargino in the vertex $t\tilde{b}\tilde{\chi}^{+}$ and neutralino in the
vertex $t\tilde{t}\tilde{\chi}^{0}$. Although generally speaking the
contribution from gluino diagram is much larger than that from the chargino
and neutralino sectors because of the strong coupling which is proportional to
$\alpha_{s} = \frac{g_{s}^{2}}{4\pi}$, the following numerical calculation
shows that there are still some occasions when the contributions from the
latter two sectors are enhanced to be comparable to the gluino contribution.
Therefore, we shall consider all the CP violating sources from the mechanism
of the MSSM in our calculation.

   According to the analysis in the MSSM theory, the CP violation in the
interactions involving top quark may be attributed to the imaginary parts
of several soft breaking parameters. The complex phases of supersymmetric
soft breaking trilinear couplings $arg(A_{t})$ and $arg(A_{b})$, enter in
the scalar quark mixing when the current eigenstates are transformed to the
mass eigenstates:
$$
\tilde{q_{L}}=(\tilde{q_{1}}\cos\theta_{q}+
              \tilde{q_{2}}\sin\theta_{q})e^{-i\phi_{q}}
$$
$$
\tilde{q_{R}}=(-\tilde{q_{1}}\sin\theta_{q}+
               \tilde{q_{2}}\cos\theta_{q})e^{i\phi_{q}}.
\eqno{(2.3)}
$$
   The complex phases of the higgsino mass parameter $\mu$ and SU(3), U(1)
gaugino mass parameters (i.e., $\phi_{\mu}$, $\phi_{SU(3)}$ and $\phi_{U(1)}$)
enter in the couplings of top quark with the gluino, chargino and neutralino
through the transformation matrices U, V, and N (see Appendix). The complex
phases of other soft breaking parameters, such as $\phi_{SU(2)}$, can be set
to be zero without loss of generality, since they can be rotated away through
suitable redefinition of the fields.

   We firstly consider the process via the collisions of polarized photon
beams with photon polarizations denoted as $\lambda_{1}$ and $\lambda_{2}$,
not measuring the spins of the final top quark pair. For the photon's
polarization vectors, we use the following formula:
$$
 \epsilon^{\mu}(p, \lambda)\epsilon^{\nu\ast}(p, \lambda')=
    \frac{\delta_{\lambda, \lambda'}}{2}
    \left(
     -g^{\mu\nu}+\frac{p^{\mu}k^{\nu}
     +p^{\nu}k^{\mu} }{(p \cdot k)}+i \lambda \epsilon^{\sigma\mu\rho\nu}
     \frac{p_{\sigma}k_{\rho} }{(p \cdot k)}
    \right),
\eqno{(2.4)}
$$
where k is an arbitrary light-like Lorentz vector. In order to describe
the CP violating effects in the process with the polarized initial states,
we introduce a CP-odd observable defined as
$$
      \xi_{CP} = \frac{\sigma_{++}-\sigma_{--}}{\sigma_{++}+\sigma_{--}},
\eqno{(2.5)}
$$
where the lower indexes appearing in the right hand of the equation denote the
helicities of the incoming photons (i.e., the $\lambda_{1}$ and $\lambda_{2}$),
$\sigma_{\pm\pm}$ represent the cross section of the process (2.1) with initial
photons being polarized and the spins of final tops being summed up. The cross
section for this process including the one-loop order supersymmetric corrections
of CP violation is expressed as
\begin{eqnarray*}
\sigma &=& <|M_{total}|^{2}> = <|M_{tree}+M^{CP}_{one-loop}|^{2}> \\
       &&\simeq <|M_{tree}|^{2}> + 2 <Re(M_{tree}^{\dag} M^{CP}_{one-loop})>.
~~~~~~~~~~~~~(2.6)
\end{eqnarray*}
Here the notation $< >$ means taking the integration over the phase space.
It is known that at the tree level the cross sections with photon helicities
$++$ and $--$ have the same value, and the interference term in Eq.(2.6) is
much smaller than the tree level term. Therefore, Eq.(2.5) can be rewritten
as following:
$$
      \xi_{CP} \simeq Re
      \left(
      \frac{<(M_{tree}^{\dag} M^{CP}_{one-loop})_{++}> -
            <(M_{tree}^{\dag} M^{CP}_{one-loop})_{--}>}
           {<|M_{tree}|^{2}_{++}>}
      \right).
\eqno{(2.7)}
$$
   Similarly, the lower indexes $\pm\pm$ appearing above represent the initial
photon helicities in the one-loop diagrams which contribute to CP violation.
Eq.(2.7) is just the practical expression used in our calculation for the
observable $\xi_{CP}$.

  Secondly, we consider the CP violating effects in the process (2.1) with
unpolarized photon beams. We assume that the final states are polarized,
and denote the spin vectors of top and anti-top quarks as $s_{1}$ and $s_{2}$,
respectively. These spin four-vectors should satisfy:
$$
      s_{i} \cdot s_{i} = -1,~~~~ s_{i} \cdot p_{i} = 0,~~~(i=1,2)
\eqno{(2.8)}
$$
which are introduced by Bjorken and Drell \cite{BD} for the massive fermion
case. In the rest frame of top quark, the spatial part of the spin vector
defined above points in the direction of top quark's spin. The wave functions
with spin vectors should satisfy the Bjorken-Drell expressions:
$$
      u(p,s)\bar{u}(p,s) = \frac{1}{2}(\rlap/p+m)(1+\gamma_{5}\rlap/s),
$$
$$
      v(p,s)\bar{v}(p,s) = \frac{1}{2}(\rlap/p-m)(1+\gamma_{5}\rlap/s).
\eqno{(2.9)}
$$
  The connection of this kind of definition with the commonly used spinor
helicity basis method is presented in Ref.\cite{Gregory}. The CP-odd
observable $\eta_{CP}$ for the unpolarized photon collisions is defined as
$$
 \eta_{CP} = \hat{p}_{1} \cdot (\vec{s}_{1} \times \vec{s}_{2}).
\eqno{(2.10.a)}
$$
  The expectation value of $\eta_{CP}$ should be
$$
  \bar{\eta}_{CP}=\frac{< \sum_{s_{1}, s_{2}}
                  \left[
                  d\sigma(s_{1}, s_{2})
                  (\hat{p}_{1} \cdot (\vec{s}_{1} \times \vec{s}_{2}))
                  \right] >} {\sigma_{total}},
\eqno{(2.10.b)}
$$
where $\hat{p}_{1}$ is the unit vector of the spatial part of $p_{1}$,
$\vec{s}_{1}$ and $\vec{s}_{2}$ are the spatial parts of the spin vectors
$s_{1}$ and $s_{2}$, respectively. The summation in the right hand of the
Eq.(2.10.b) should be performed over all the possible spins of top quark pair.
Note that the component of top(anti-top) quark's spin vector along its momentum
has no contribution to the triple product observable $\eta_{CP}$,
thus we only consider the top(anti-top) quark spin with its spatial component
in the direction perpendicular to vector $\vec{p}_{1}$.
In order to satisfy Eq.(2.8), we choose the time parts of $s_{1}$ and
$s_{2}$ to be zero and $\vec{s}_{1}$ and $\vec{s}_{2}$ to be unit vectors
lying on the plane perpendicular to $\vec{p}_{1}$($\vec{p}_{2}$).
If we define $\theta_{1}(\theta_{2})$ as the angle between $\vec{s}_{1}
(\vec{s}_{2})$ and the reaction plane, our calculation shows that the
interference term appearing in Eq.(2.6) can be expressed as the function of
$\cos{\theta_{1}}$, $\cos{\theta_{2}}$, $\sin{\theta_{1}}$ and
$\sin{\theta_{2}}$:
\begin{eqnarray*}
  2 Re (M_{tree}^{\dag}(s_{1}, s_{2})
       M^{CP}_{one-loop}(s_{1}, s_{2})) &=&
    C_{0}+C_{1}\cos{\theta_{1}}+C_{2}\cos{\theta_{2}}+ \\
 && C_{3}\sin{\theta_{1}}+ C_{4}\sin{\theta_{2}}+ \\
 && C_{5}\cos{\theta_{1}}\cos{\theta_{2}}+ \\
 && C_{6}\sin{\theta_{1}}\sin{\theta_{2}}+ \\
 && C_{7}\cos{\theta_{1}}\sin{\theta_{2}}+ \\
 && C_{8}\sin{\theta_{1}}\cos{\theta_{2}},
~~~~~~~~~~~~(2.11)
\end{eqnarray*}

   To evaluate the observable $\bar{\eta}_{CP}$, here we introduce Cartesian
coordinate frame(x, y, z) in the CMS of this reaction. In this frame, $\hat{z}$
is a unit vector along the outgoing direction of top quark, whereas $\hat{x}$
is defined in the production plane of top quark pair, and both $\hat{x}$ and
$\hat{y}$ are located in the plane perpendicular to the outgoing direction of
top quark. Then all the orthogonal combinations of the top and anti-top's spins,
which have non-zero contributions to CP-odd observable $\bar{\eta}_{CP}$, are
listed below:

$$ (1)~~s_{1}=(0,+\hat{x})=(0,+1,0,0),~~~~
        s_{2}=(0,+\hat{y})=(0,0,+1,0);$$
$$ (2)~~s_{1}=(0,+\hat{x})=(0,+1,0,0),~~~~
        s_{2}=(0,-\hat{y})=(0,0,-1,0);$$
$$ (3)~~s_{1}=(0,-\hat{x})=(0,-1,0,0),~~~~
        s_{2}=(0,+\hat{y})=(0,0,+1,0);$$
$$ (4)~~s_{1}=(0,-\hat{x})=(0,-1,0,0),~~~~
        s_{2}=(0,-\hat{y})=(0,0,-1,0);$$
$$ (5)~~s_{1}=(0,+\hat{y})=(0,0,+1,0),~~~~
        s_{2}=(0,+\hat{x})=(0,+1,0,0);$$
$$ (6)~~s_{1}=(0,+\hat{y})=(0,0,+1,0),~~~~
        s_{2}=(0,-\hat{x})=(0,-1,0,0);$$
$$ (7)~~s_{1}=(0,-\hat{y})=(0,0,-1,0),~~~~
        s_{2}=(0,+\hat{x})=(0,+1,0,0);$$
$$ (8)~~s_{1}=(0,-\hat{y})=(0,0,-1,0),~~~~
        s_{2}=(0,-\hat{x})=(0,-1,0,0).
\eqno{(2.12)}
$$
Because that the contribution from tree-level cross section to Eq.(2.10.b)
turns out to be zero, only the interference term between the tree level
amplitude and the CP violating one-loop amplitudes should be evaluated.
By using Eq.(2.11) and Eq.(2.12), the observable $\bar{\eta}_{CP}$ can be
worked out as
$$
\bar{\eta}_{CP} = \frac{< \sum_{s_{1}, s_{2}}
                  \left[
                  2 Re(M_{tree}^{\dag}M^{CP}_{one-loop})
                  (\hat{p}_{1} \cdot (\vec{s}_{1} \times \vec{s}_{2}))
                  \right] >} {\sigma_{total}}
                = \frac{4 <C_{8} - C_{7}>}{\sigma_{total}}.
\eqno{(2.13)}
$$
   In our calculation, the dimensional reduction method and the
on-mass-shell(OMS) renormalization scheme are adopted to eliminate the
ultraviolet divergences appearing at one-loop order\cite{Denner}. The
detailed steps
and formula of renormalization can be referred to Ref.\cite{han}.
The explicit evaluation demonstrate that the vertex and self-energy
diagrams shown in Fig.1 have no contribution to $\xi_{CP}$, but
contribute to $\bar{\eta}_{CP}$, when the CP-violating phases are not
all zero.

\vskip 5mm
\noindent{\large\bf III. Numerical Calculations and Discussions}

        The two CP-odd observables defined above are strongly related with
the CP violating parameters involved in the Yukawa couplings with top quark,
i.e., $V_{t\tilde{t}\tilde{g}}$, $V_{t\tilde{b}\tilde{\chi}^{+}}$,
and $V_{t\tilde{t}\tilde{\chi}^{0}}$. The corresponding Lagrangians of the
interactions can be written as

\begin{eqnarray*}
        {\cal L}_{\bar{t}_{\alpha}\tilde{t}_{\beta}\tilde{g}_{\gamma}} &=&
              -\sqrt{2} g_{s} T_{\alpha\beta}^{\gamma} \bar{t}_{\alpha}
              (P_{R} \tilde{t}_{L}^{\beta}  - P_{L} \tilde{t}_{R}^{\beta} )
               \tilde{g}_{\gamma}e^{i\phi_{SU(3)}} + h.c. \\
        {\cal L}_{\bar{t}\tilde{t}\tilde{\chi}_{i}^{0}} &=&
              -\sqrt{2} g \bar{t} \tilde{t}_{L} \tilde{\chi}_{i}^{0}
              \left\{
                \frac{m_{t}}{2 m_{W} \sin\beta} P_{L} N_{i4}^{\ast} +
                (\frac{1}{6} \tan\theta_{W} N_{i1} + \frac{1}{2} N_{i2}) P_{R}
              \right\} \\
           && +\sqrt{2} g \bar{t} \tilde{t}_{R} \tilde{\chi}_{i}^{0}
              \left\{
                -\frac{m_{t}}{2 m_{W} \sin\beta} P_{R} N_{i4} +
                 \frac{2}{3} \tan\theta_{W} P_{L} N_{i1}^{\ast}
              \right\}
              + h.c. \\
        {\cal L}_{\bar{t}\tilde{b}\tilde{\chi}_{i}^{+}} &=&
              -g \bar{t} \tilde{b}_{L} P_{R} U_{i1} \tilde{\chi}_{i}^{+} +
              \frac{g m_{b}}{\sqrt{2} m_{W} \cos\beta}
                    \bar{t} \tilde{b}_{R} P_{R} U_{i2} \tilde{\chi}_{i}^{+} \\
           && + \frac{g m_{t}}{\sqrt{2} m_{W} \sin\beta}
                    \bar{t} \tilde{b}_{L} P_{L} V_{i2}^{\ast}
                    \tilde{\chi}_{i}^{+}
              + h.c.,
~~~~~~~~~~~(3.1)
\end{eqnarray*}
where $T_{\alpha\beta}^{\gamma}$ are the SU(3) generators,
$\alpha, \beta, \gamma$ are the color indices, $g_{s}$ and $g$ are the
strong and weak coupling constant, and $P_{R,L}=(1\pm\gamma_{5})/2$.
It can be seen that the squark mixing angles $\theta_{\tilde{q}}$ and
phases $\phi_{\tilde{q}}~~(\tilde{q}=\tilde{t}, \tilde{b})$ are involved
in the couplings when we express the Lagrangian with the mass eigenstates
$\tilde{q}_{1}, \tilde{q}_{2}$ instead of the weak eigenstates
$\tilde{q}_{L}, \tilde{q}_{R}$. Because normally the CP effects from the
gluino sector is much more important than from the chargino and neutralino
sectors, the most considerable contribution of the squark mixing phases
to CP violation is the effect of the phase angle $\phi_{\tilde{t}}$ on the
vertex of top-stop-gluino.

      The CP violating effects may be induced by the complex phases of $SU(3)$,
$U(1)$ mass parameters and Higgs mass parameter $\mu$, wherein $\phi_{SU(3)}$
only emerges in the gluino sector through the Majorana mass term \cite{Dugan},
$\phi_{U(1)}$ is only involved in the neutralino sector through the
diagonalizing matrix N, and $\phi_{\mu}$ impresses both neutralino and chargino
sectors through matrices U, V and N. It should be noticed that these phases
take parts in the CP violation not only through the chargino and neutralino
mass transformation matrices, but also through their mass spectra. The chargino
and neutralino masses may vary by about one half of their original values when
these complex phases are varied.

      Although there are some constraints on the supersymmetric parameters
in the theory, such as grand unification(GUT), in the following analysis
we do not put any extra limitations on them for the general discussion.
In the numerical calculation, we assume the following corresponding input
parameters by default, in case that no special declaration has been presented
on them:

$$
      \sqrt{\hat{s}}=600~GeV,~~~ \tan\beta=2,~~~ m_{\tilde{g}}=150~GeV,
$$
$$
      m_{\tilde{t}_{1}}=150~GeV,~~ m_{\tilde{t}_{2}}=400~GeV,~~
      m_{\tilde{b}_{1}}=270~GeV,~~ m_{\tilde{b}_{2}}=280~GeV,
$$
$$
      \theta_{\tilde{t}}=\theta_{\tilde{b}}=\pi/6,~~
      \phi_{\tilde{t}}=\phi_{\tilde{b}}=\pi/5,
$$
$$
      |M_{U(1)}|=320~GeV,~~~ M_{SU(2)}=250~GeV,~~~ |\mu|=220~GeV,
$$
$$
      \phi_{U(1)}=\pi/4,~~~ \phi_{SU(3)}=0,~~~ \phi_{\mu}=4\pi/3,
\eqno{(3.2)}
$$

  Our calculation shows that $\phi_{SU(3)}$ has no contribution to the cross
section and CP-odd observables for our specific process, and both the
CP-odd observables $\xi_{CP}$ and
$\bar{\eta}_{CP}$ vanish when all the complex phases $\phi_{\tilde{t}},~
\phi_{\tilde{b}},~\phi_{U(1)}$, and $\phi_{\mu}$ are set to
zero. The CP-odd observables $\xi_{CP}$ and $\bar{\eta}_{CP}$ as the
functions of $\phi_{U(1)}$ and $\phi_{\mu}$ are plotted in Fig.2 and Fig.3,
respectively. The curves in these two figures show that both the two
observables are $2 \pi$ periodic odd functions of $\phi_{U(1)}$ and
$\phi_{\mu}$, i.e., they satisfy
$\xi_{CP}(\pi-\phi_{U(1),\mu})=-\xi_{CP}(\pi+\phi_{U(1),\mu})$ and
$\bar{\eta}_{CP}(\pi-\phi_{U(1),\mu})=-\bar{\eta}_{CP}(\pi+\phi_{U(1),\mu})$,
when the corresponding other phases are all neglected. However,
they do not have the strict symmetries for transformation $\phi_{U(1)}
\rightarrow \pi-\phi_{U(1)}$ or $\phi_{\mu} \rightarrow \pi-\phi_{\mu}$.
Fig.4 plots the CP-odd observables as functions of $\phi_{\tilde{t}}$,
where the other phases are zero, because the strength of top-stop-gluino
coupling depends on the value of $\phi_{\tilde{t}}$, and comparatively
the contributions from squark mixing phases to non-QCD sectors are small
such that they can be omitted in general cases. From the curve we can see
that the CP violation is maximized when $\phi_{\tilde{t}}= \frac{\pi}{4}$,
the absolute values of $\xi_{CP}$ and $\bar{\eta}_{CP}$ being over one
percent, which is much larger than in the usual cases.

    In Fig.5(a,b,c,d) we depict the dependences of the CP-violating
parameters $\xi_{CP}$ and $\bar{\eta}_{CP}$ on the c.m. energy $\sqrt{\hat{s}}$,
with the contributions from gluino, chargino, neutralino and overall diagrams,
respectively. The threshold effect near the energy region
$\sqrt{\hat{s}}=2 m_{t}$ can be seen obviously in these figures.
From the figure 5(a) and 5(c) it is shown that at the position of
$\sqrt{\hat{s}} \sim 2 m_{\tilde{t}_{2}}=800~GeV$ there is an abrupt
turning-point in $\xi_{CP}$ and a spike in $\bar{\eta}_{CP}$ in each of the
curves, which are caused by the resonance effects mainly coming from the
triangle diagram including the coupling $\gamma\tilde{t}\bar{\tilde{t}}$,
which occurs only in gluino and neutralino sectors. Fig.5 (b) shows that the
curve of $\bar{\eta}_{CP}$ contributed by chargino sector reaches its maximal
and minimal values at the positions of
$\sqrt{\hat{s}} \sim 2 m_{\tilde{\chi}^{+}_{1,2}} = 400, 580~GeV$,
respectively, due to the resonance effects happening in the coupling
$\gamma\tilde{\chi}^{+} \tilde{\chi}^{-}$. However, the curve of $\xi_{CP}$
has an almost plain maximum region in the range of 400 GeV to 580 GeV. It is
because of the mergence of several individual resonance effects originating
from the diagrams of chargino sector at the positions of
$\sqrt{\hat{s}} \sim 2 m_{\tilde{b}_{1,2}},~2 m_{\tilde{\chi}^{+}_{1,2}}$,
where $m_{\tilde{b}_{1,2}} \sim 270,~280~GeV$, $m_{\tilde{\chi}^{+}_{1,2}}=
200,~290~GeV$.
That leads to the CP-violation parameter $\bar{\eta}_{CP}$ contributed from
all sectors has two peaks in Fig.5(d), one is around $\sqrt{\hat{s}} \sim
800~GeV$ and another is about $\sqrt{\hat{s}} \sim  500~GeV$. From the
magnitude order of the curves we may infer that the chargino loop diagrams
impress the CP-odd observable quite considerably and in some region its
CP-violating effect may be comparable to the contribution from gluino sector,
whereas that from neutralino is always the smallest among the three sectors.
In addition, Fig.5 also shows that the absolute values of both the two CP-odd
observables approach to zero with increasing $\sqrt{\hat{s}}$, when the c.m.s.
energy of photons is beyond $800~GeV$.

     We vary the parameter $\tan{\beta}$ from 0.5 to 100 with other parameters
being taken as in Eq.(3.2). The results are depicted in Fig.6, among which
(a) stands for the chargino sector contribution and (b) for contribution from
neutralino sector. We find that the absolute values of both the two observables
decrease steadily with increasing $\beta$, except at the end of each curve of
Fig.6(a) where $\beta>1.4$. And generally the contributed parts of $\xi_{CP}$
($\bar{\eta}_{CP}$) from the chargino and neutralino sectors have opposite
signs, so that they cancel with each other to some extent. In Fig.7 and Fig.8,
the two observables as the functions of $|M_{SU(2)}|$ and $|\mu|$ are plotted,
respectively. The curves in both figures have the similar property. In both
the two figures it is shown that the CP violating effect is quite weak when
$|M_{SU(2)}|$ or $|\mu|$ is near zero, because that in these cases the mass
of the lightest chargino or neutralino is very small and thus the contributions
from the CP-violating phases are suppressed. We can also see from the figure
that each curve has a sharp slope at the position of $|M_{SU(2)}| \sim 260~GeV$
(Fig.7) or $|\mu| \sim 230~GeV$ (Fig.8) due to the resonance effects when
$\sqrt{\hat{s}}=600~GeV \sim 2 m_{\tilde{\chi}^{+}_{2}}$. And when the
absolute value of $M_{SU(2)}$ or $|\mu|$ becomes larger, the CP-violating
effect gets lower steadily. Fig.9 plots the CP-violating parameters as
functions of $|M_{U(1)}|$, which reflects the impression of U(1) mass parameter
on the CP violation through neutralino diagonalizing matrix N and its mass
spectra. The curves go down steadily except that in the region of
$M_{U(1)} < 60~GeV$, $\xi_{CP}$ rises sharply. In Fig.10 we only consider the
contribution from QCD sector and plot the dependence of CP violation on the
mass of gluino($m_{\tilde{g}}=|M_{SU(3)}|$). From the figure one can see the
curves rise sharply with increasing gluino mass at the low $|M_{SU(3)}|$
region, then go down steadily as $|M_{SU(3)}|$ becomes larger. The maximal
values are $1.15\%$ for $\xi_{CP}$ and $0.98\%$ for $\bar{\eta}_{CP}$ around
the position of $M_{SU(3)} \sim 200~GeV$.

\vskip 10mm
\noindent
{\Large{\bf IV.Summary}}
\vskip 5mm
  In this work we have studied all the contributions to the CP-odd observables
in the process $\gamma \gamma \rightarrow t \bar{t}$ in the frame of the
MSSM with complex soft breaking SUSY parameters. The CP violating effects in
this process are related to the complex phases of $\mu$, $A_{t}$, $A_{b}$,
$M_{SU(3)}$ and $M_{U(1)}$ through the diagonalization of the complex stop,
sbottom, chargino and neutralino mass matrices. We introduce the CP-odd
observables $\xi_{CP}$ and $\bar{\eta}_{CP}$ to describe the CP violating
effects in polarized and unpolarized photon collision cases, respectively.
Our calculation shows that they can be different from zero and are typically
of the order of $10^{-4} \sim 10^{-2}$, if CP violation really exists. The CP
violating effect contributed by gluino sector is generally the most important,
whereas neutralino and chargino exchanges in the loop diagrams play less
important roles, but cannot be neglected in some cases. We find that it is
possible to observe the CP violating effects in top quark pair production via
polarized and unpolarized photon fusions by using optimal observables and
favorable parameters. Therefore, probing CP violation in this process is a
rather prospective goal for future photon-photon colliders.

\vskip 5mm
\noindent{\large\bf Acknowledgement:}
These work was supported in part by the National Natural Science
Foundation of China(project numbers: 19675033) and the Youth Science
Foundation of the University of Science and Technology of China.

\vskip 5mm
\noindent{\large\bf Appendix: Transformation Matrices U,V and N}

      One can get the physical mass spectra and transformation matrices of
charged gauginos when diagonalizing the following mass matrix:
$$
      X = \left(
      \begin{array}{ll}
         M_{SU(2)} & m_{W}\sqrt{2}\sin\beta  \\
         m_{W}\sqrt{2}\cos\beta &  |\mu|e^{i\phi_{\mu}}
      \end{array}
      \right),
\eqno{(A.1)}
$$
where the complex phase of $M_{SU(2)}$ has been neglected because it is
a trivial one. The two $2 \times 2$ unitary matrices U, V are defined to
diagonalize the matrix $X$, namely,
$$
      U^{\ast}XV^{\dag} = X_{D},
\eqno{(A.2)}
$$
where $X_{D}$ is a diagonal matrix with non-negative entries. The two
diagonal elements of this matrix are worked out in general case as
\begin{eqnarray*}
      M_{\pm}^{2} &=& \frac{1}{2}
        \left\{
         M_{SU(2)}^{2}  + |\mu|^{2} + 2 m_{W}^{2} \pm
        \left[ (M_{SU(2)}^{2}-|\mu|^{2})^{2} + 4 m_{W}^{4} \cos^{2}2\beta +
        \right. \right. \\
&& \left. \left. 4 m_{W}^{2} (M_{SU(2)}^{2} +
         |\mu|^{2}+ 2 M_{SU(2)} |\mu|
         \sin2\beta \cos\phi_{\mu})
                     \right]^{1/2}
                \right\},
~~~~~~~~~~~~~~~~~~(A.3)
\end{eqnarray*}
which just stand for the masses of chargino $\tilde{\chi}^{+}_{1}$ and
$\tilde{\chi}^{+}_{2}$. The diagonalizing matrices U and V have very
complicated forms depending on the complex phase of $\mu$. In general,
we can write
$$
      U = \left(
      \begin{array}{ll}
         \cos\theta_{U}e^{i(\phi_{1}+\xi_{1})} &
         \sin\theta_{U}e^{i(\phi_{1}+\xi_{1}+\delta_{U})} \\
        -\sin\theta_{U}e^{i(\phi_{2}+\xi_{2}-\delta_{U})} &
         \cos\theta_{U}e^{i(\phi_{2}+\xi_{2})}
      \end{array}
      \right)
$$
$$
      V = \left(
      \begin{array}{ll}
         \cos\theta_{V}e^{i(\phi_{1}-\xi_{1})} &
         \sin\theta_{V}e^{i(\phi_{1}-\xi_{1}+\delta_{V})} \\
        -\sin\theta_{V}e^{i(\phi_{2}-\xi_{2}-\delta_{V})} &
         \cos\theta_{V}e^{i(\phi_{2}-\xi_{2})}
      \end{array}
      \right),
\eqno{(A.4)}
$$

      where the $\xi_{1}$ and $\xi_{2}$ can be any arbitrarily chosen phases.
It indicates the matrices U and V satisfying Eq.(A.2) are not unique, namely,
some arbitrary phases may be introduced into the physical fields. But our
calculation shows that they have no effects on the CP-odd observables.
The explicit forms of the other constant angles and phases depending on the
input parameters are given as
$$
      \tan{\theta_{U}}=\sqrt{\frac{M_{+}^{2}-M_{SU(2)}^{2}-2m_{W}^{2}\sin^{2}\beta}
                               {M_{+}^{2}-|\mu|^{2}-2m_{W}^{2}\cos^{2}\beta}},
$$
$$
      \tan{\theta_{V}}=\sqrt{\frac{M_{+}^{2}-M_{SU(2)}^{2}-2m_{W}^{2}\cos^{2}\beta}
                               {M_{+}^{2}-|\mu|^{2}-2m_{W}^{2}\sin^{2}\beta}},
$$
$$
      e^{i2\phi_{1}}=\frac{\cos\theta_{U}}{\cos\theta_{V}} \cdot
                     \frac{M_{+}^{2}+M_{SU(2)}|\mu|\tan\beta e^{i\phi_{\mu}}
                                   -2m_{W}^{2}\sin^{2}\beta}
                         {M_{+}(M_{SU(2)}+|\mu| \tan\beta e^{i\phi_{\mu}})},
$$
$$
      e^{i2\phi_{2}}=\frac{\cos\theta_{V}}{\cos\theta_{U}} \cdot
                     \frac{M_{-}^{2}+M_{SU(2)}|\mu|\tan\beta e^{i\phi_{\mu}}
                                   -2m_{W}^{2}\sin^{2}\beta}
                         {M_{-}(M_{SU(2)}\tan\beta+|\mu| e^{-i\phi_{\mu}})},
$$
$$
      e^{i\delta_{U}}=\frac{M_{SU(2)}+|\mu| e^{i\phi_{\mu}}\tan\beta}
                           {|M_{SU(2)}+|\mu| e^{i\phi_{\mu}}\tan\beta|},
$$
$$
      e^{i\delta_{V}}=\frac{M_{SU(2)}\tan\beta+|\mu| e^{i\phi_{\mu}}}
                           {|M_{SU(2)}\tan\beta+|\mu| e^{i\phi_{\mu}}|},
\eqno{(A.5)}
$$
where $M_{\pm}$ can be evaluated from Eq.(A.3).

     As for the case of neutral gauginos, we can obtain the transformation
matrix $N$ by diagonalizing the following $4 \times 4$ mass matrix:
$$
      Y = \left(
      \begin{array}{llll}
         |M_{U(1)}|e^{i\phi_{U(1)}} & 0 & -m_{Z}\sin\theta_{W}\cos\beta
                                      & m_{Z}\sin\theta_{W}\sin\beta  \\
         0 & M_{SU(2)} & m_{Z}\cos\theta_{W}\cos\beta
                                      & -m_{Z}\cos\theta_{W}\sin\beta  \\
         -m_{Z}\sin\theta_{W}\cos\beta & m_{Z}\cos\theta_{W}\cos\beta
                                      & 0 & -|\mu|e^{i\phi_{\mu}}      \\
         m_{Z}\sin\theta_{W}\sin\beta & -m_{Z}\cos\theta_{W}\sin\beta
                                      & -|\mu|e^{i\phi_{\mu}} & 0
      \end{array}
      \right).
\eqno{(A.6)}
$$
      Again the parameter $M_{SU(2)}$ can be set to be real. The
transformation matrix $N$ is chosen such that
$$
      N^{\ast}YN^{\dag} = Y_{D}
\eqno{(A.7)}
$$
and should be unitary. $Y_{D}$ is a $4 \times 4$ diagonal matrix with four
non-negative entries. To obtain the mass spectra $Y_{D}$ and the transformation
matrix $N$, we separate $Y$ into the real part $Y_{1}$ and the imaginary part
$Y_{2}~~(Y=Y_{1}+iY_{2})$, and define an $8\times8$ matrix as
$$
      Y' = \left(
      \begin{array}{ll}
                       Y_{1}  &  -Y_{2} \\
                      -Y_{2}  &  -Y_{1}
      \end{array}
      \right).
\eqno{(A.8)}
$$
      We have proved that, if the eight eigenvalues of $Y'$ can be worked out,
the four positive ones among them will just be the four entries of $Y_{D}$,
i.e., the physical masses of neutralino. Meanwhile, if the four corresponding
eigenvectors of $Y'$ to the four positive eigenvalues are denoted as
$(R_{i1}~R_{i2}~R_{i3}~R_{i4}~I_{i1}~I_{i2}~I_{i3}~I_{i4})^{T}$,
where $i$ is from 1 to 4, the transformation matrix $N$ will take the form of
$$
      N = \left(
      \begin{array}{llll}
         R_{11}-iI_{11} & R_{12}-iI_{12} & R_{13}-iI_{13} & R_{14}-iI_{14} \\
         R_{21}-iI_{21} & R_{22}-iI_{22} & R_{23}-iI_{23} & R_{24}-iI_{24} \\
         R_{31}-iI_{31} & R_{32}-iI_{32} & R_{33}-iI_{33} & R_{34}-iI_{34} \\
         R_{41}-iI_{41} & R_{42}-iI_{42} & R_{43}-iI_{43} & R_{44}-iI_{44}
      \end{array}
      \right).
\eqno{(A.9)}
$$

\vskip 20mm

\vskip 20mm
\noindent{\Large\bf Figure captions}
\vskip 5mm

\noindent
{\bf Fig.1} The Feynman diagrams at tree level and the MSSM one-loop order
        diagrams contributing to the CP violation for process $\gamma\gamma
        \rightarrow t\bar{t}$. (a) tree level diagram; (b)-(e) vertex diagrams;
        (f)-(h) box diagrams; (i) quartic coupling diagram, and (j) self-energy
        diagrams. The $\tilde{q}$ and $\tilde{s}$ in Fig.1 respectively denote
        the following corresponding particles: In figures (b), (c), (f) and (h),
        $\tilde{q}= \tilde{b}_{1,2},~\tilde{s}= \tilde{\chi}^{+}_{1,2}$,
        whereas in diagrams (d), (e), (g), (i) and (j), there are three sets
        of combinations: $\tilde{q}= \tilde{t}_{1,2},~\tilde{s}=\tilde{g};~~~
        \tilde{q}=\tilde{b}_{1,2}, ~\tilde{s}=\tilde{\chi}^{+}_{1,2};~~~$
        and $\tilde{q}=\tilde{t}_{1,2}, ~\tilde{s}=\tilde{\chi}^{0}_{1,2,3,4}$.
        The diagrams with incoming photons exchanged are not shown in the
        figures except for (i).

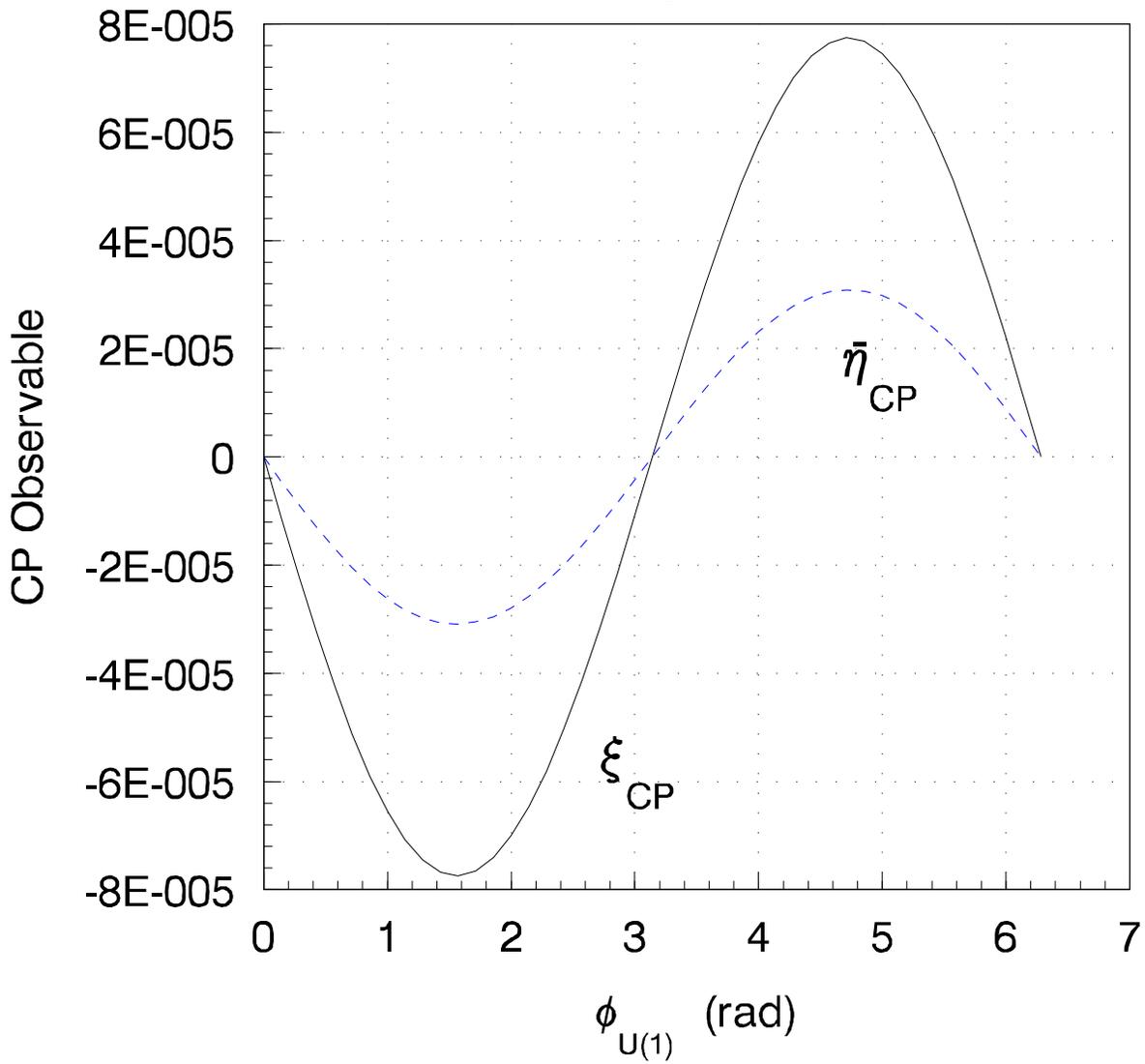
{\bf Fig.2} The CP-violating parameters as the functions of $\phi_{U(1)}$
        with $\phi_{\tilde{t},\tilde{b}}=\phi_{\mu}=\phi_{SU(3)}=0$.
        The solid line is for $\xi_{CP}$ and the dashed line is for
        $\bar{\eta}_{CP}$.

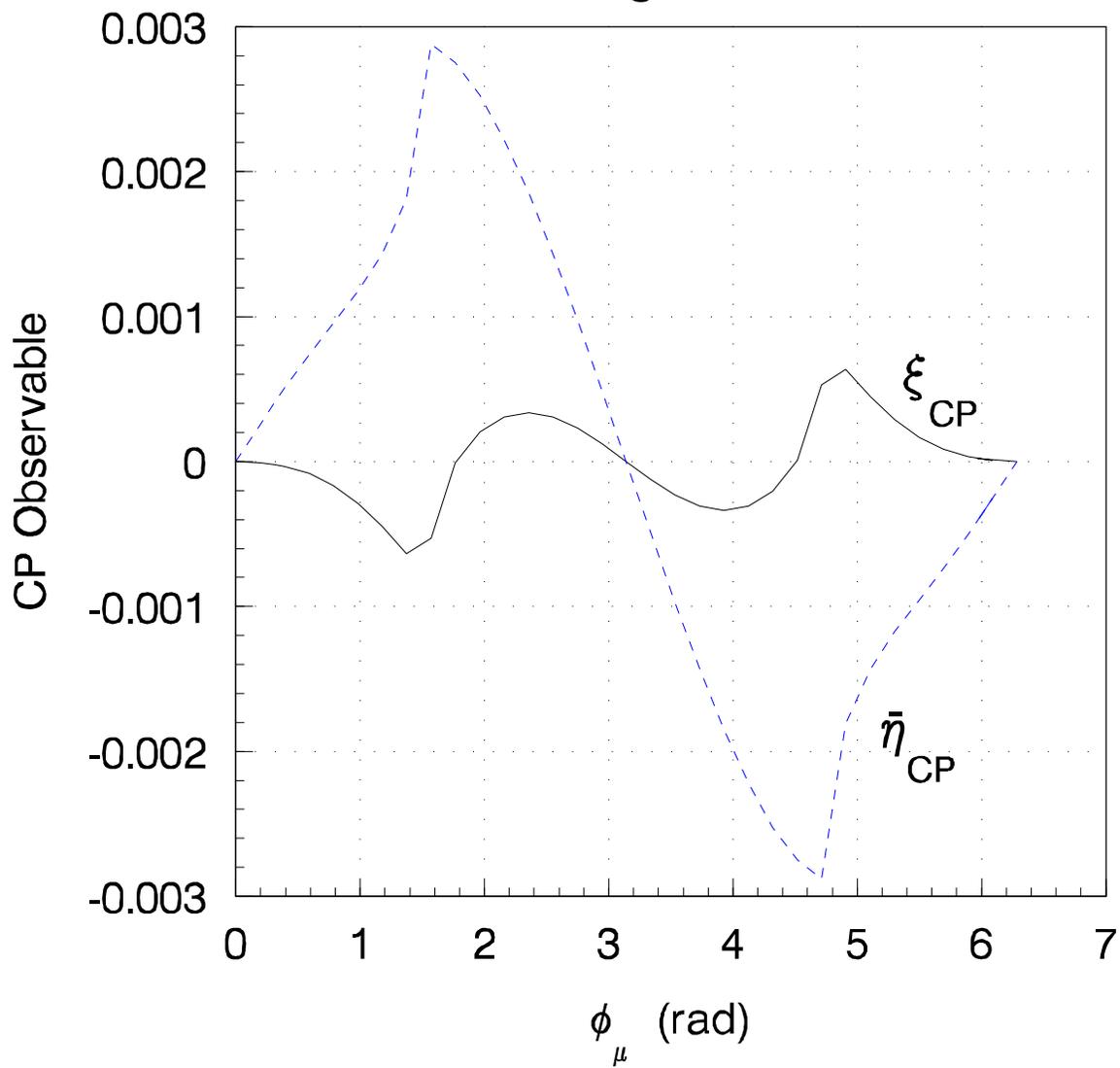
{\bf Fig.3} The CP-violating parameters as the functions of $\phi_{\mu}$
        with $\phi_{\tilde{t},\tilde{b}}=\phi_{U(1)}=\phi_{SU(3)}=0$.
        The solid line is for $\xi_{CP}$ and the dashed line is for
        $\bar{\eta}_{CP}$.

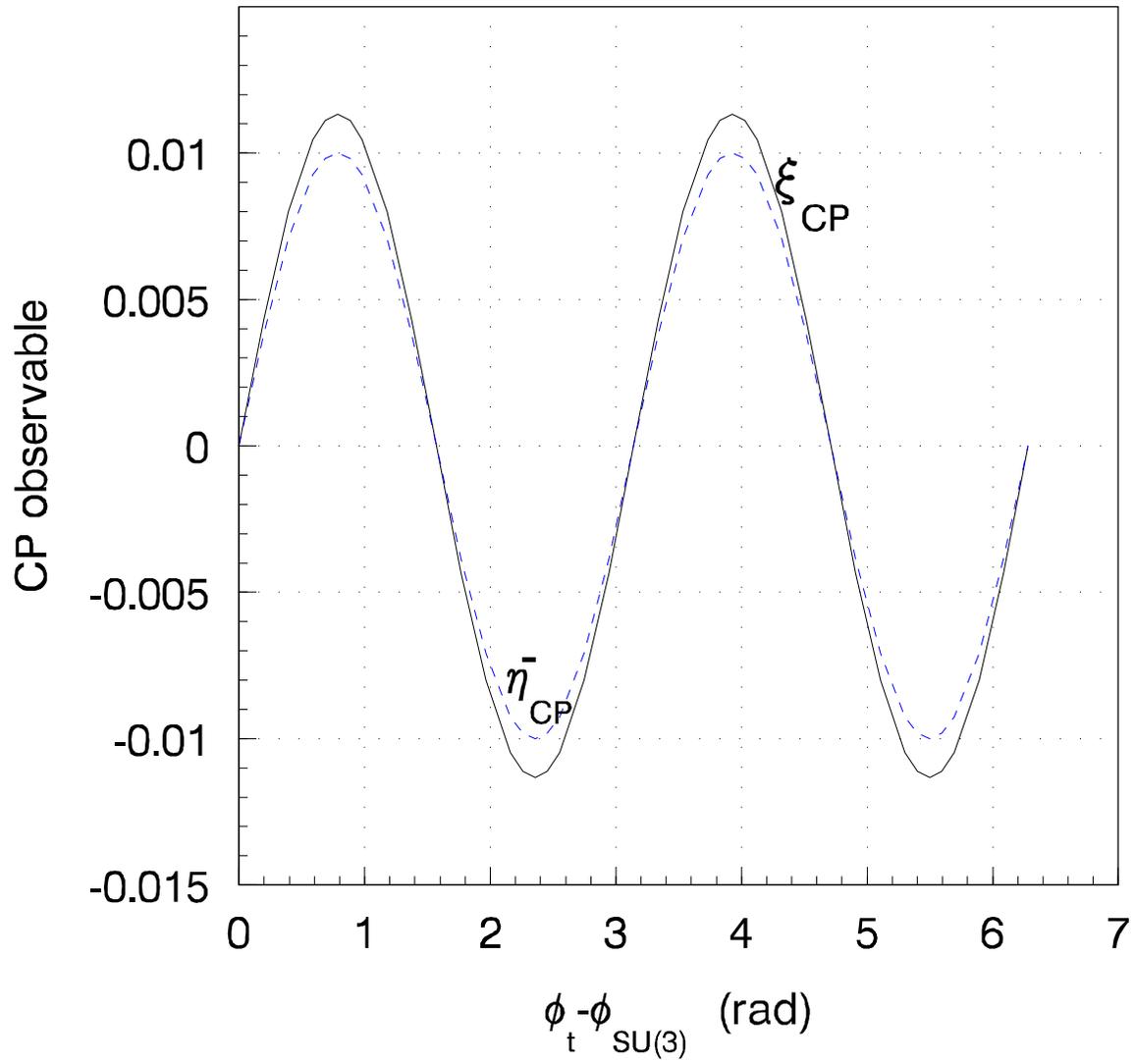
{\bf Fig.4} The CP-violating parameters as the functions of
        $\phi_{\tilde{t}}-\phi_{SU(3)}$ with $\phi_{\mu}=\phi_{U(1)}=0$.
        The solid line is for $\xi_{CP}$ and the dashed line is for
        $\bar{\eta}_{CP}$.

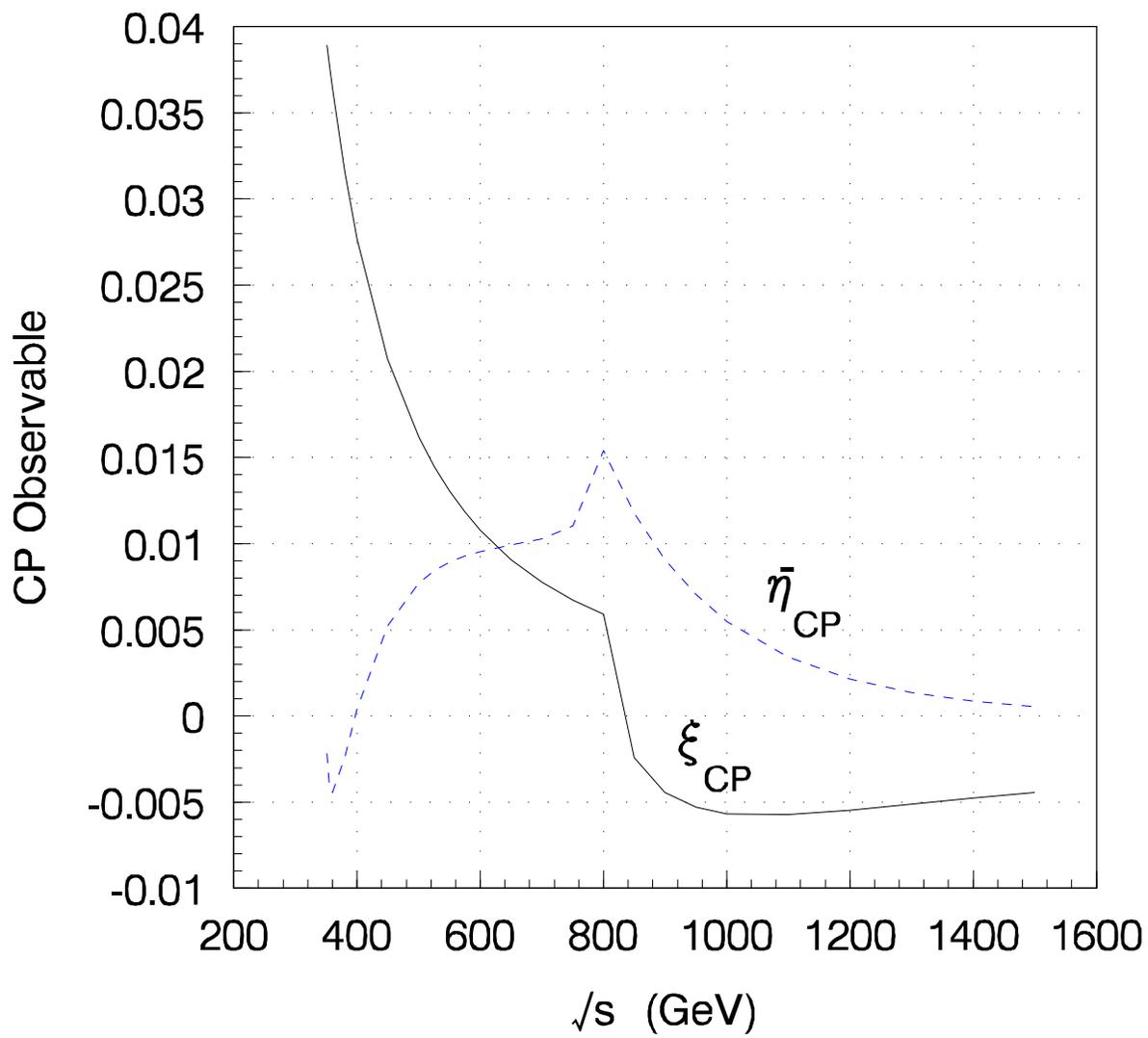
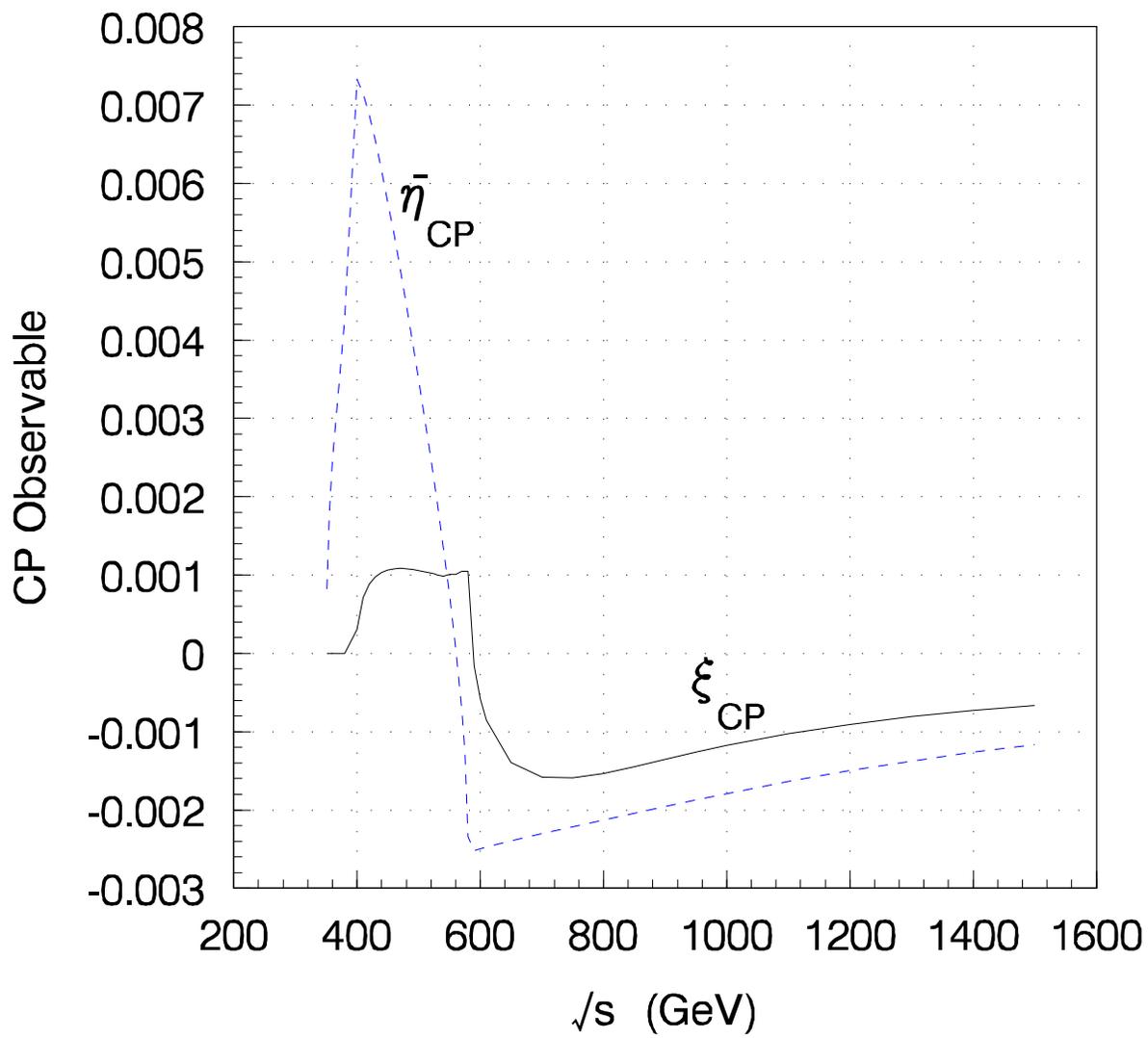
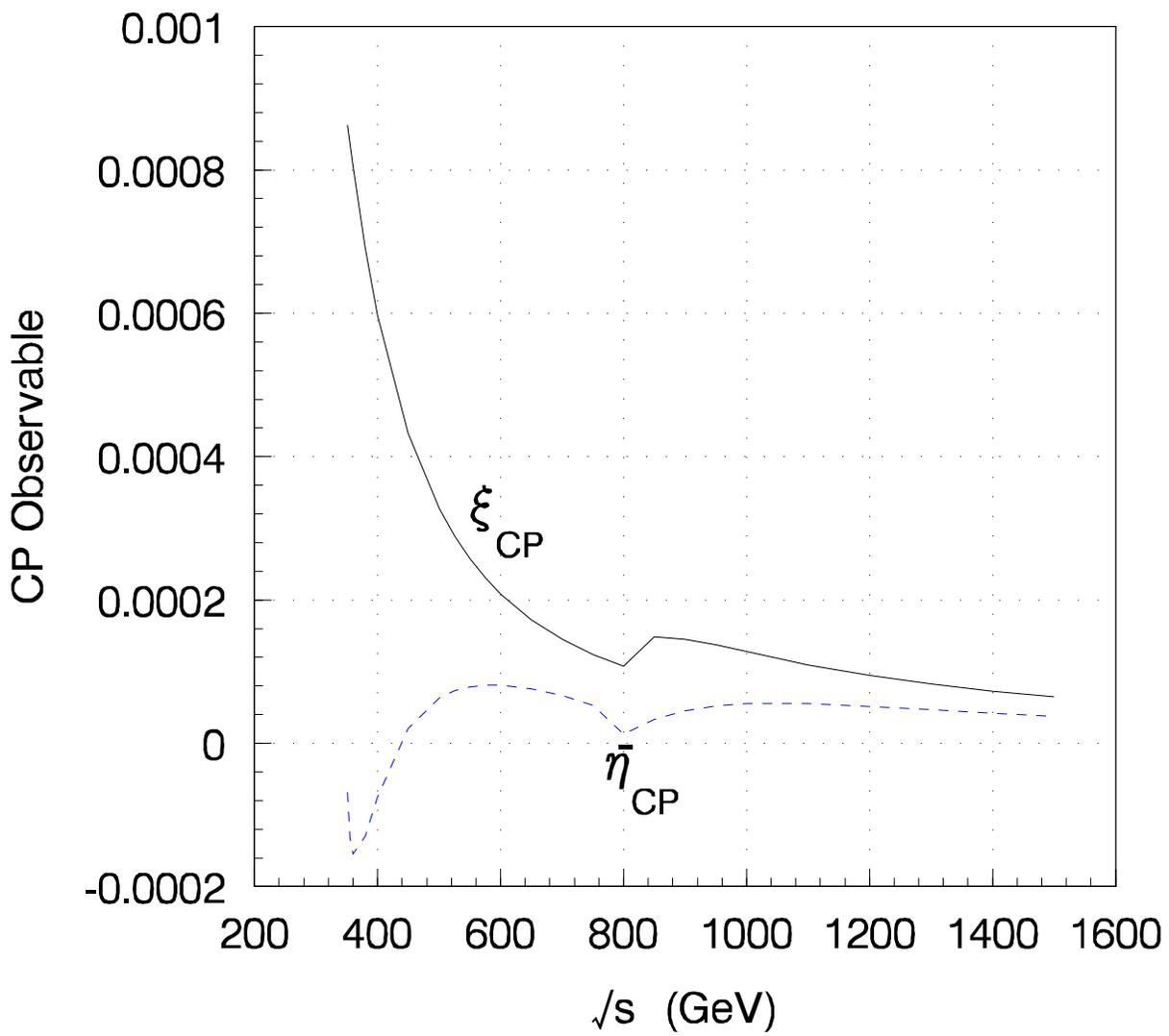
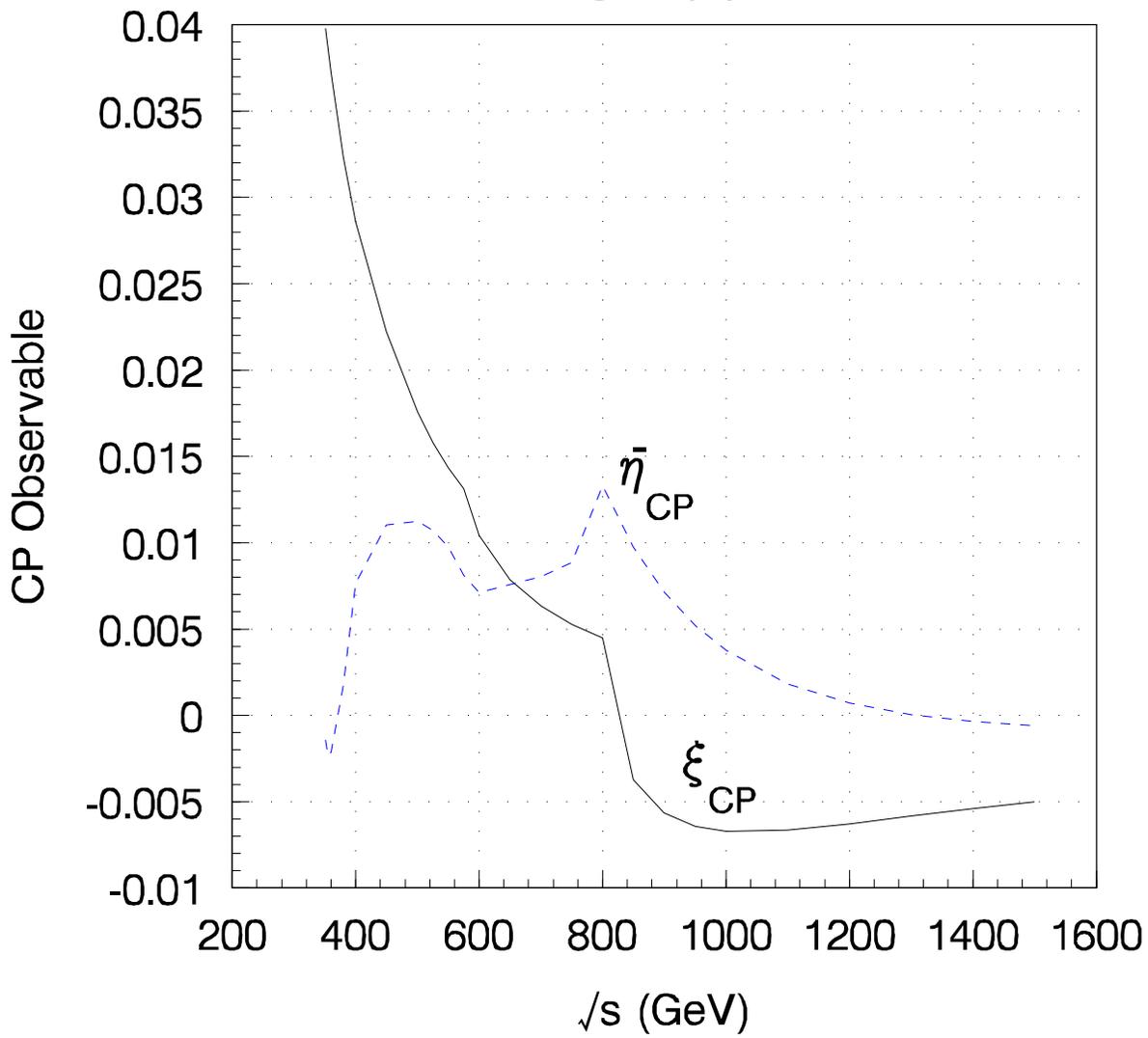
{\bf Fig.5} The CP-violating parameters as the functions of $\sqrt{\hat{s}}$
        with the values of other parameters shown in Eq.(3.2).
        (a) the gluino sector only. (b) the chargino sector only.
        (c) the neutralino sector only. (d) all three sectors together.
        The solid line is for $\xi_{CP}$ and the dashed line is for
        $\bar{\eta}_{CP}$.

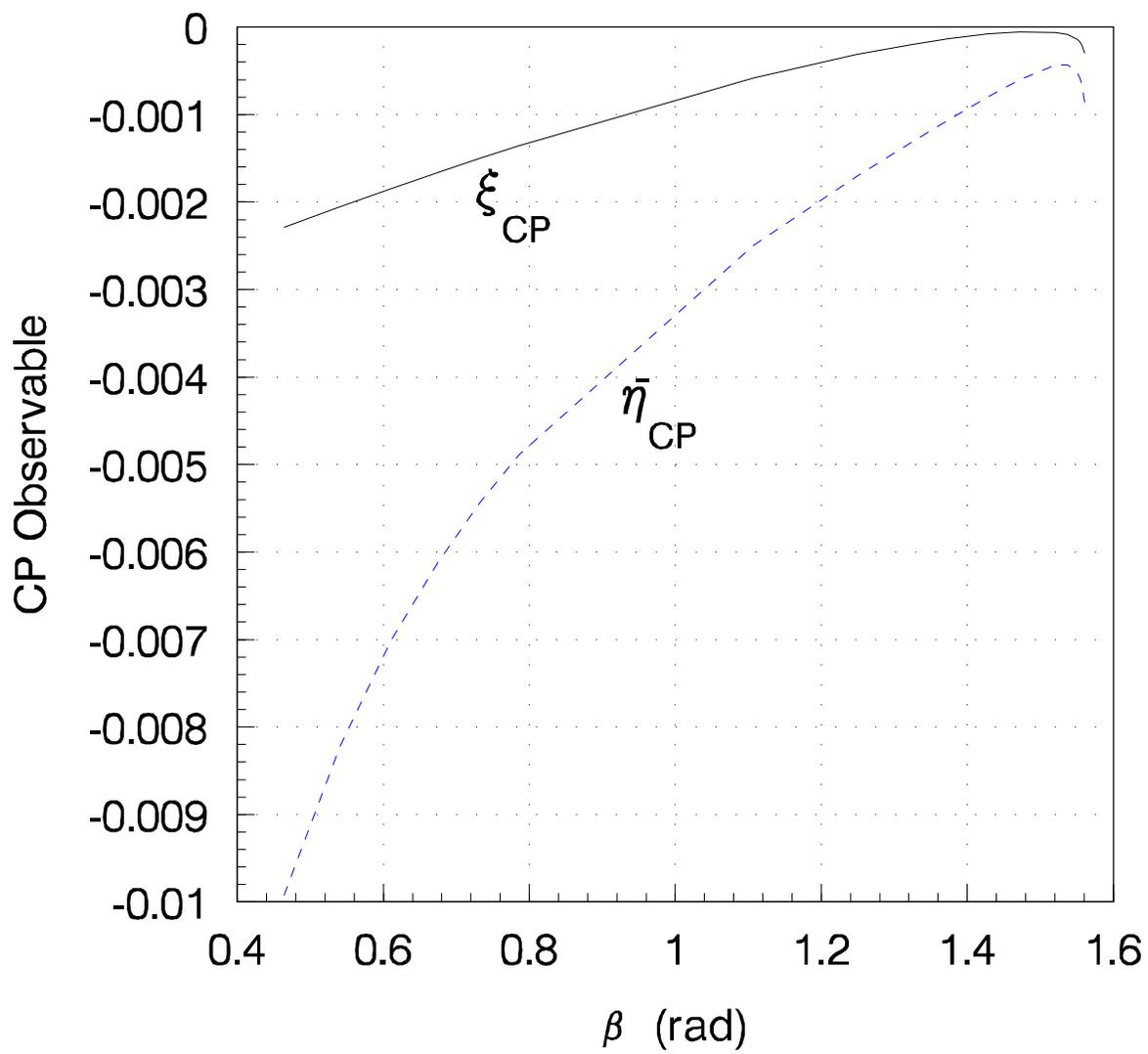
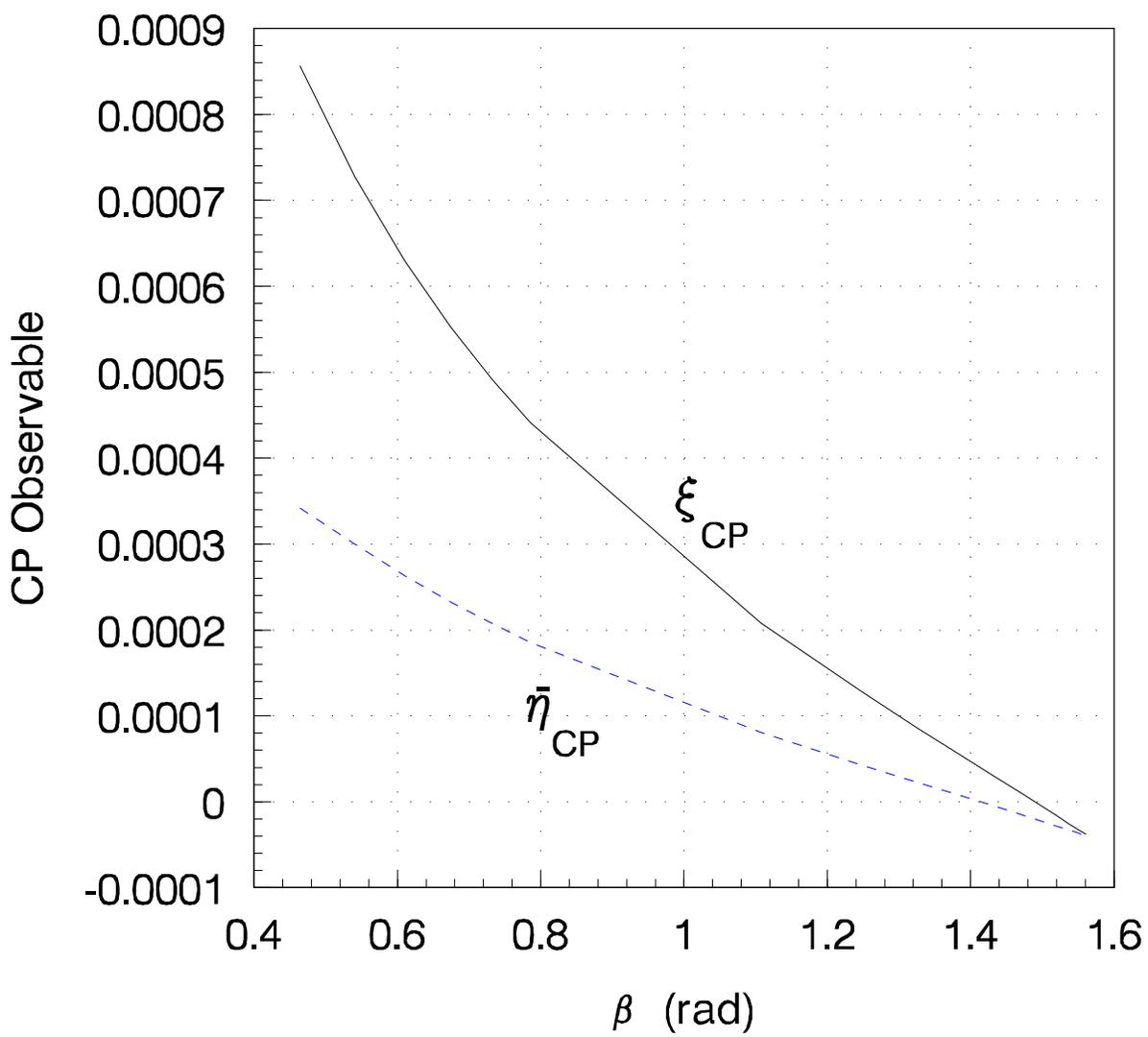
{\bf Fig.6} The CP-violating parameters as the functions of $\beta$
        with the values of other parameters shown in Eq.(3.2).
        (a) the chargino sector only. (b) the neutralino sector only.
        The solid line is for $\xi_{CP}$ and the dashed line is for
        $\bar{\eta}_{CP}$.

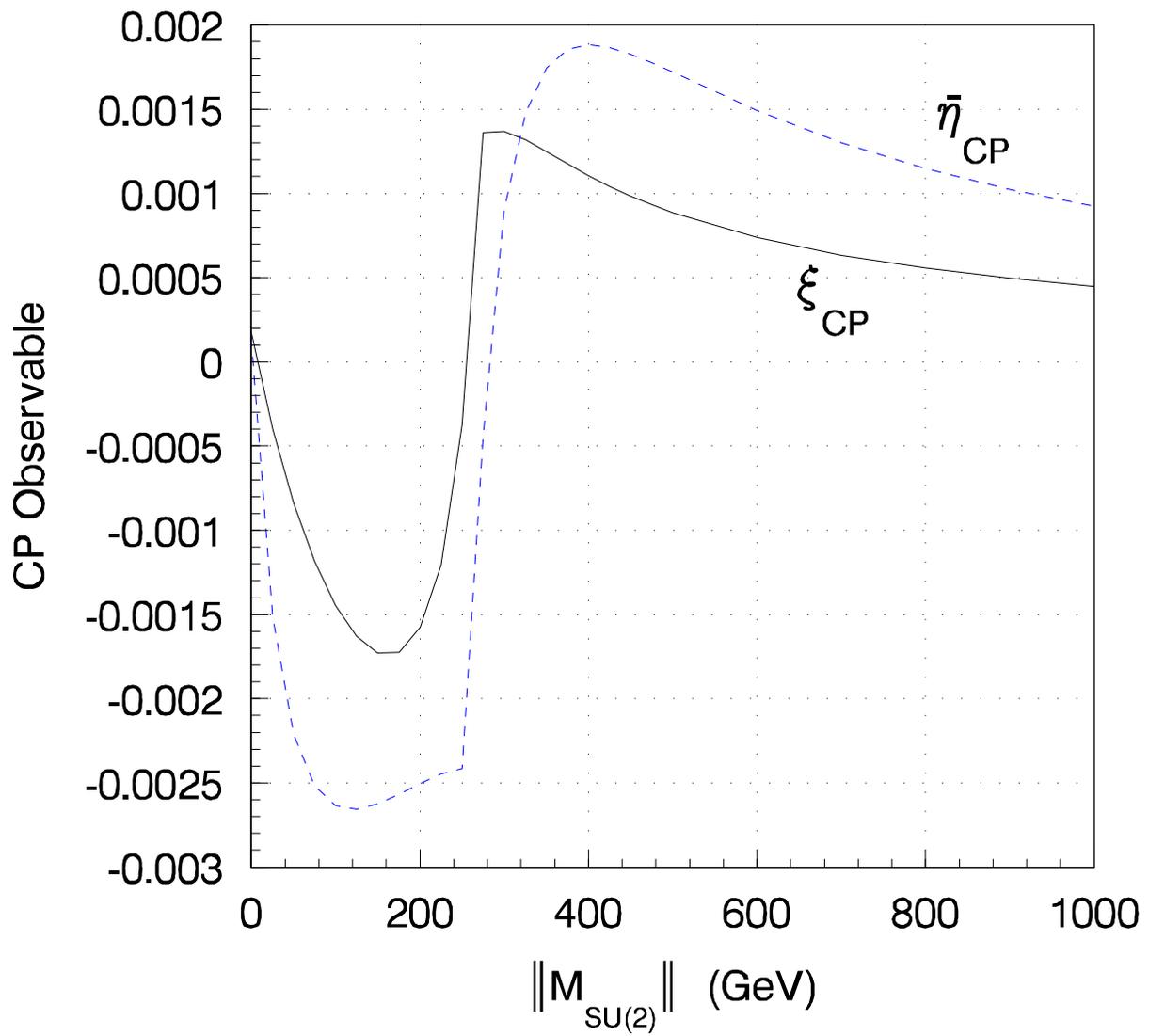
{\bf Fig.7} The CP-violating parameters contributed by chargino and
        neutralino sectors as the functions of $M_{SU(2)}$
        with $\phi_{\tilde{t}}=\frac{\pi}{5}$ and the
        values of other parameters shown in Eq.(3.2).
        The solid line is for $\xi_{CP}$ and the dashed line is for
        $\bar{\eta}_{CP}$.

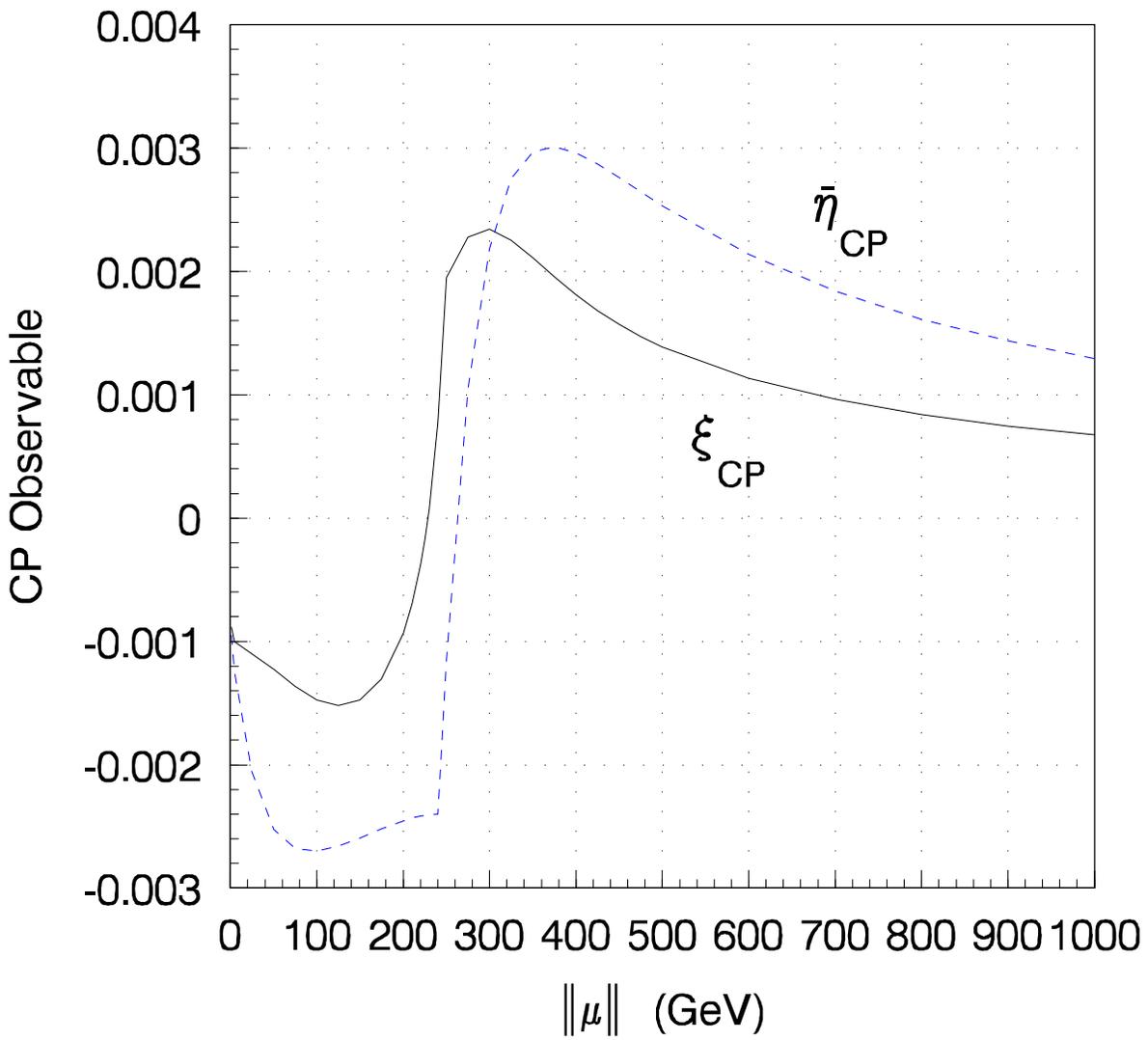
{\bf Fig.8} The CP-violating parameters contributed by chargino and
        neutralino sectors as the functions of $|\mu|$
        with $\phi_{\tilde{t}}=\frac{\pi}{5}$ and the
        values of other parameters shown in Eq.(3.2). The solid line is
        for $\xi_{CP}$ and the dashed line is for $\bar{\eta}_{CP}$.

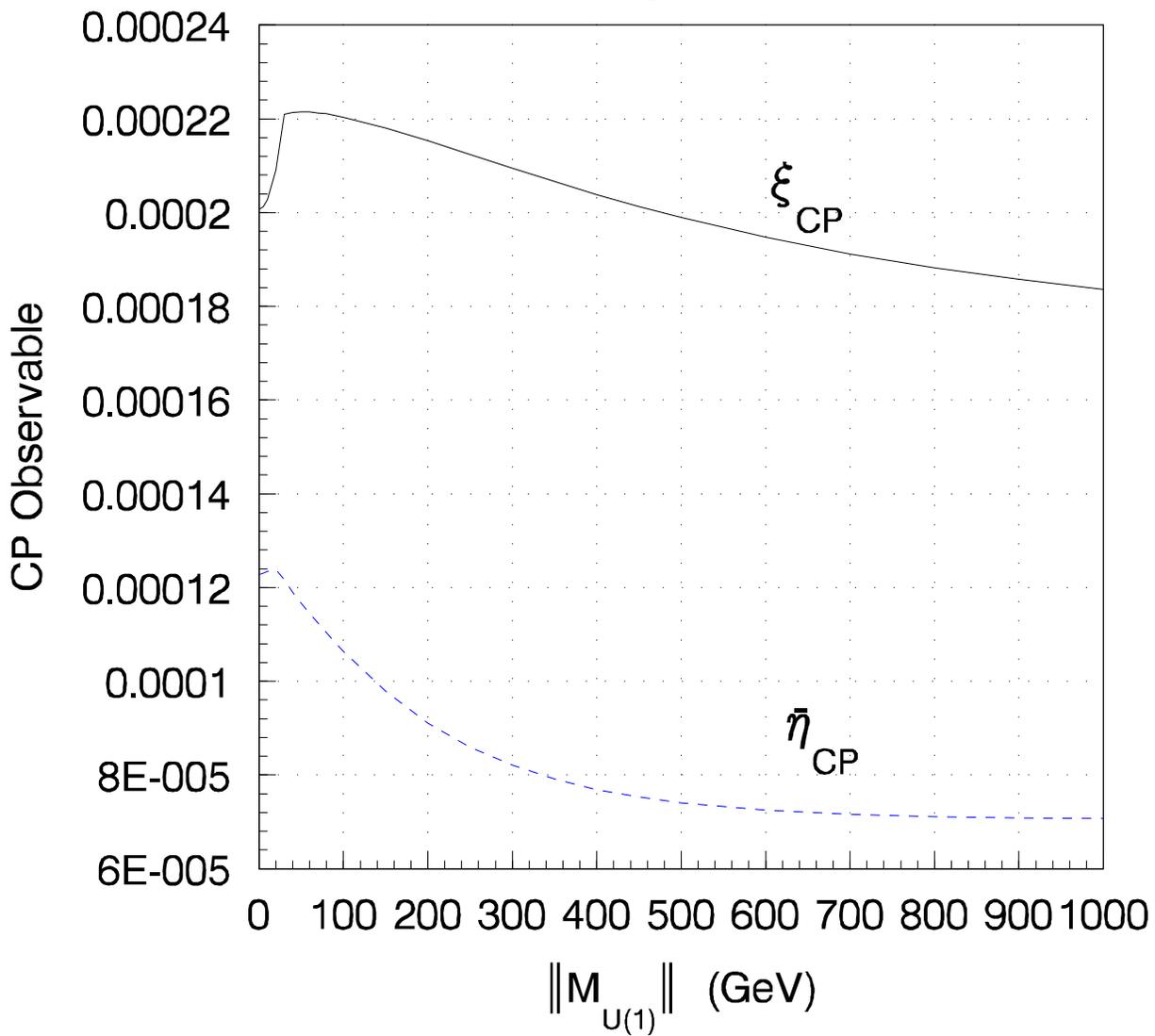
{\bf Fig.9} The CP-violating parameters as the functions of $M_{U(1)}$,
        only the neutralino sector being taken into consideration.
        The values of other parameters are taken as shown in Eq.(3.2).
        The solid line is for $\xi_{CP}$ and the dashed line is for
        $\bar{\eta}_{CP}$.

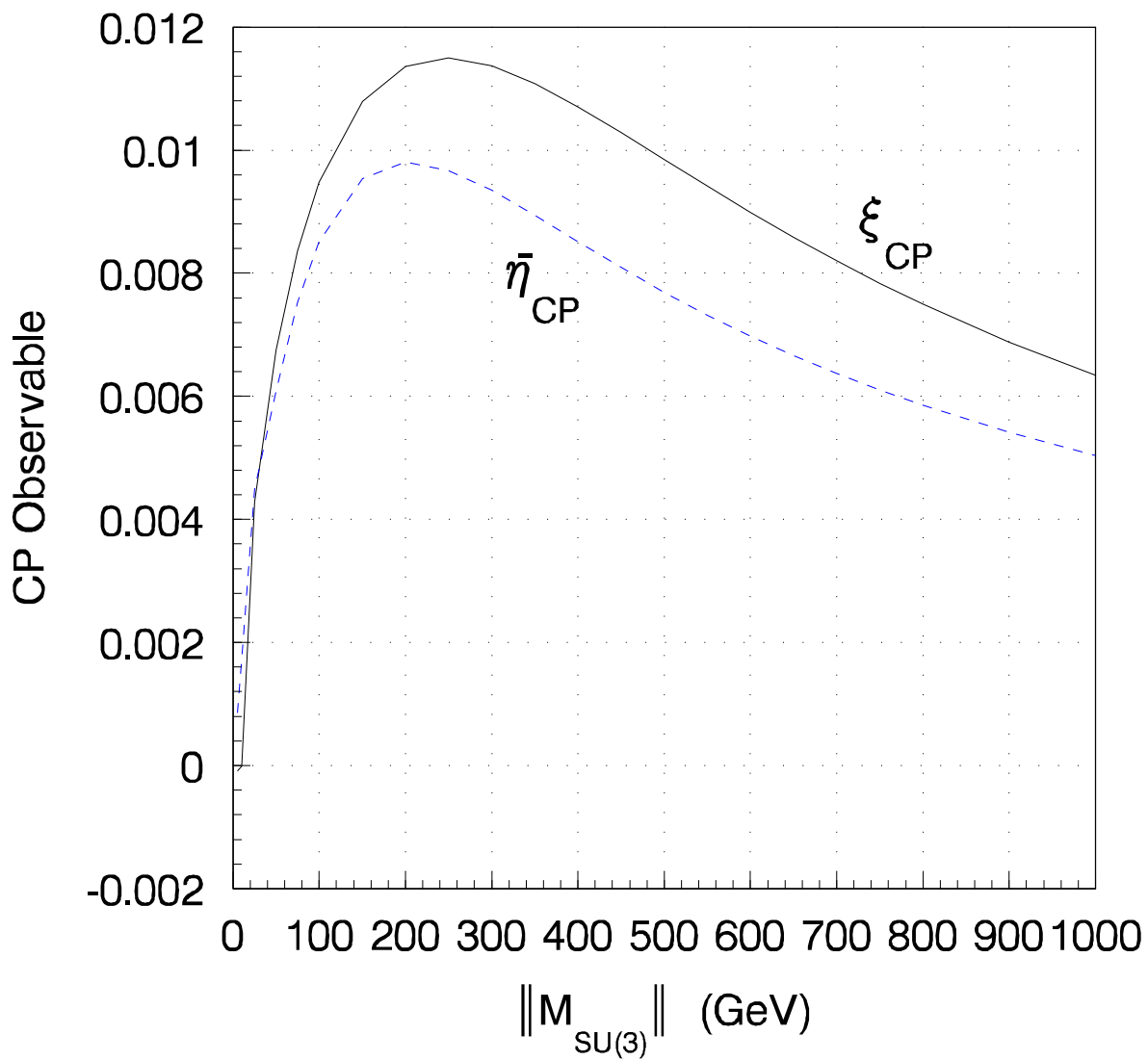
{\bf Fig.10} The CP-violating parameters as the functions of $M_{SU(3)}$
        by taking $\phi_{\mu}=\phi_{U(1)}=0$. The values of
        other parameters are taken as shown in Eq.(3.2). The solid line is
        for $\xi_{CP}$ and the dashed line is for $\bar{\eta}_{CP}$.
\vskip 3mm
\noindent

\vskip 3mm
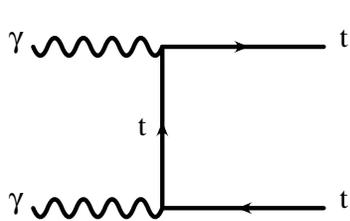
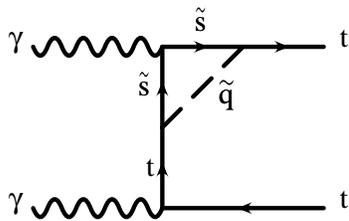
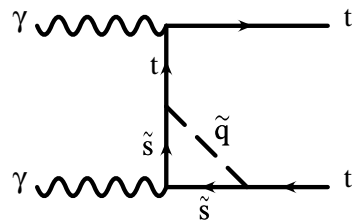
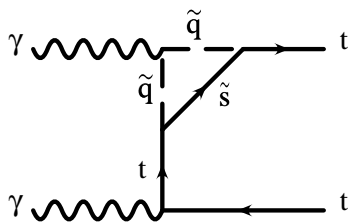
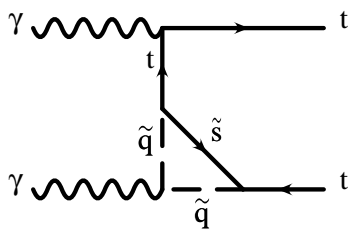
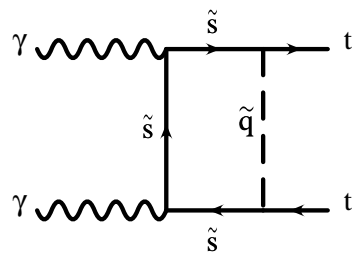
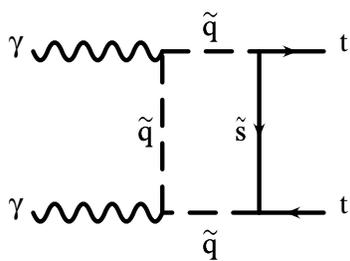
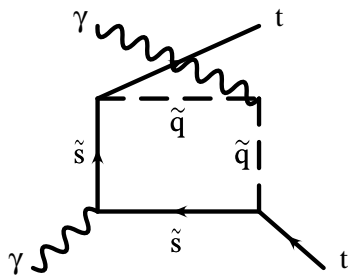
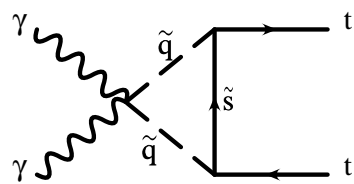
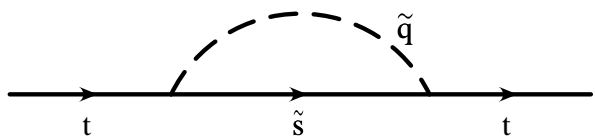
\end{document}